\documentclass[11pt]{article}
\usepackage{hyperref}
\begin{document}

\title{Splash control using geometric targets}

\author{G. Juarez, Y. Zhang, T. Gastopoulos, and P. E. Arratia\\
\\
\vspace{6pt}
Department of Mechanical Engineering and Applied Mechanics,\\
University of Pennsylvania, Philadelphia, PA 19104, USA}

\maketitle

\begin{abstract}
Experiments of water droplets impacting small geometric posts of equal dimension 
to the drop diameter are shown in this fluid dynamics video. High speed photography 
shows that the dynamics of drop splashing are significantly affected by the geometrical 
boundary of the target and that finger formation and drop break up are accurately controlled.
\end{abstract}

\section*{Introduction}

When a liquid drop impacts with a smooth solid dry surface, it will expand horizontally
outward until internal and external forces create asymmetry. This instability leads to 
the droplet breaking apart in a ``splash''. The dynamics of splashing as a result of liquid 
drops and solid surface impacts have been studied extensively, particularly when the surface
is much larger than the drop. Here, this fluid dynamics video investigates impacts between 
fluid droplets and geometrical substrates of equal length scale.

The liquid used has a  viscosity of 1 cP and surface tension of 0.353 N/m. It is composed 
of de-ionized water and food coloring, for image enhancement purposes. Liquid drops are 
created by allowing water at room temperature ($T = 25 \ ^{\circ}$C) to slowly drip out of 
a small capillary tube. Water is injected into the tube using a low-noise syringe pump 
(Harvard Instruments). Reproducible drops are created with average diameter of $2.8$ mm. 
Drops fall from a height of $13$ cm before striking the target, hitting the surface with 
a velocity of $2$ m/s resulting in a Reynolds (Re) number equal to $540$ and a Weber 
(We) number equal to $250$. Here, Re is defined as $\rho v d/\mu$ and We is defined as 
$\rho v^2 d/\gamma$ where $\rho$ is the fluid density, $v$ is the drop velocity, $d$ is 
the drop diameter, $\mu$ is dynamic viscosity , and $\gamma$ is the surface tension. Top 
and side view images are recorded using high speed photography at 30,000 frames per second 
(Fastcam SA1.1, Photron USA Inc.). The images in this fluid dynamics video are shown at 25 
frames per second.

The targets are posts made out of polyoxymethylene plastic with no surface treatments. The cylindrical 
post has a diameter of $2.7$ mm and a surface area of $5.73$ mm$^2$. The other polygonal posts
have cross sectional shapes such as an equilateral triangle, a square, a pentagon, and a 
hexagon with equal impacting surface area of $5.73$ mm$^2$.

When a droplet impacts a cylindrical post, the resulting splash is axisymmetric. However,
as the geometry of the post is changed, the dynamics of the splash and finger formation
is altered drastically. Surprisingly, the resulting splash resembles the shape of the 
target with a clockwise rotation equal to $\pi/N$, where $N$ is the number of vertices. 
The break up of the drop and number of fingers that form can be accurately controlled 
and is observed to be equal to the number of vertices. For example, a drop impacting with 
a triangular post results in a triangular splash that is shifted by $60^\circ$ with 
respect to the triangular post. This splash then breaks up into three distinct fingers
after reaching its maximum splash radius.

The video showing drops impacting small geometric targets as described above can
be found at the following URLs:
\begin{itemize}
\item \href{http://ecommons.library.cornell.edu/bitstream/1813/}{High-resolution}
\item \href{http://ecommons.library.cornell.edu/bitstream/1813/}{Low-resolution}
\end{itemize}
This video has been submitted to the American Physical Society division of 
fluid dynamics annual meeting showcase, the \emph{Gallery of Fluid Motion 2010}.

\end{document}